\documentclass[article,twocolumn,aps,superscriptaddress,longbibliography]{revtex4-1}

\usepackage{graphicx}
\usepackage{dcolumn}% Align table columns on decimal point
\usepackage{bm}% bold math
\usepackage{epstopdf}
\usepackage{float}
\usepackage{amsmath}
\usepackage{xcolor}
\usepackage{comment}
\usepackage[normalem]{ulem}
\graphicspath{{./images/}}
\usepackage{pgfplots}

\usepackage{pgfplots}\pgfplotsset{compat=1.8}
\usepackage{lineno}
\usepackage{cancel}

\usepackage{tikz}
\usetikzlibrary{calc}

\usepackage{xcolor}
\definecolor{lightblue}{rgb}{0.2,0.2, 0.9}

\begin{document}
\title{{Lattice Polaron in a Bose-Einstein Condensate of Hard-Core Bosons }}
\author{Moroni Santiago-Garc\'ia}
\affiliation{Instituto Nacional de Astrof\'isica, \'Optica y Electr\'onica,
Calle Luis Enrique Erro No.1 Santa Mar\'ia Tonantzintla, Puebla CP 72840, Mexico}
\author{Shunashi G. Castillo-L\'opez}
\author{Arturo Camacho-Guardian}
\email{acamacho@fisica.unam.mx}
\affiliation{Instituto de F\'isica, Universidad Nacional Aut\'onoma de M\'exico, Apartado Postal 20-364, Ciudad de M\'exico C.P. 01000, Mexico}

%\linenumbers
\date{\today}
\begin{abstract}
Lattice polarons, quasiparticles arising from the interaction between an impurity and its surrounding bosonic environment confined to a lattice system, have emerged as a platform for generating complex few-body states, probing many-body phenomena, and addressing long-standing problems in physics. In this study, we employ a variational ansatz to investigate the quasiparticle and spectral properties of an impurity coupled to a condensate gas of hard-core bosons in a two-dimensional optical lattice. Our findings demonstrate that the polaron features can be tuned by adjusting the filling factor of the bath, revealing intriguing polaron characteristics in the strongly interacting regime. These results offer valuable insights for lattice polaron experiments with ultracold gases and can serve as a guide for new experiments in emergent quantum devices, such as moiré materials, where optical excitations can be described in terms of hard-core bosons.

\end{abstract}

\maketitle

\section{Introduction}

Ultracold gases have served as a robust platform for quantum simulation of exotic many-body physics~\cite{bloch2012quantum,Georgescu2014,gross2017quantum,schafer2020tools}. 
This permitted the realization of quantum analogs of well-understood phenomena~\cite{greiner2002quantum,tarruell2018quantum}, as well as the exploration of physics beyond the accessible regimes in condensed matter physics. 
Nowadays, the versatility of these systems allows for addressing concise proposals for long-standing open problems such as high-Tc superconductivity.  
Bose polaron physics in ultracold gases has attracted much attention given this context, and it has dramatically stimulated new theoretical and numerical approaches~\cite{rath2013field,Ardila2015impurity,Shchadilova2016,grusdt2018strong,Christensen2015,christianen2023phase,levinsen2021quantum,levinsen2015impurity,Sun2017,Sun2017b,christianen2022chemistry,christianen2022bose,Yoshida2018,Massignan2021,Yegovtsev2022,atoms9020022,cayla2023observation,nielsen2019critical,Guenther2021,bighin2022impurity,camacho2023polaritons,Liu2024,isaule2024functional} to understanding the experimental realization of this phenomenon far beyond its original formulation~\cite{jorgensen2016observation,hu2016bose,ardila2019analyzing,yan2020bose,skou2021non,Skou2022,morgen2023quantum,etrych2024universal}.

In optical lattices, impurity physics has renewed interest in probing the Mott-insulator to superfluid transition~\cite{colussi2023lattice}, topological phases~\cite{grusdt2016interferometric,Camacho2019,Pimenov2021}, band geometry~\cite{pimenov2024polaron}, magnetic polarons~\cite{koepsell2021microscopic,Grusdt2018,Wang2021,Nielsen2021,Nielsen2022}, few-body physics~\cite{Ding2022a}, non-equilibrium dynamics~\cite{isaule2024bound}, and polaron physics in strongly correlated Fermi-Hubbard models~\cite{amelio2024polaron}. The study of lattice polarons is further motivated by the advances in quantum gas microscopy, which enables the imaging of individual atoms~\cite{bakr2009quantum,bakr2010probing}, providing intricate spatial details of quantum states that complement traditional spectroscopic information~\cite{koepsell2021microscopic,hilker2017revealing}.

On the other hand, recent developments with two-dimensional van der Waals heterostructures place quantum gases into new territories where polaron physics and Bose-Fermi mixtures are being realized with control and tunability~\cite{ruiz2023bose} with excitons and charge carriers (electrons and holes).  Polaron physics has demonstrated to be a powerful tool to sense correlated phases of matter~\cite{Shimazaki2021,smolenski2021signatures,Mazza2022,Amelio2023,julku2023exciton}. In these experiments, Bose-Fermi Hubbard systems may arise as a consequence of an emergent moiré potential~\cite{kennes2021moire,huang2022excitons,tang2020simulation}. In multilayers, spatially indirect excitons (electron and hole sitting in different layers) can arise, imprinting strong dipole-dipole interactions, which can effectively lead to a Hubbard model sensitive to the intrinsic quasi-bosonic character of the excitations~\cite{huang2023non}. Indeed, for relevant experiments, the excitons in a moiré superlattice may behave as hard-core bosons, which have been predicted to exhibit superfluid~\cite{Remez2022} and supersolid phases~\cite{julku2022nonlocal}. 
\begin{figure}[t]
\centering
\includegraphics[width=1\columnwidth]{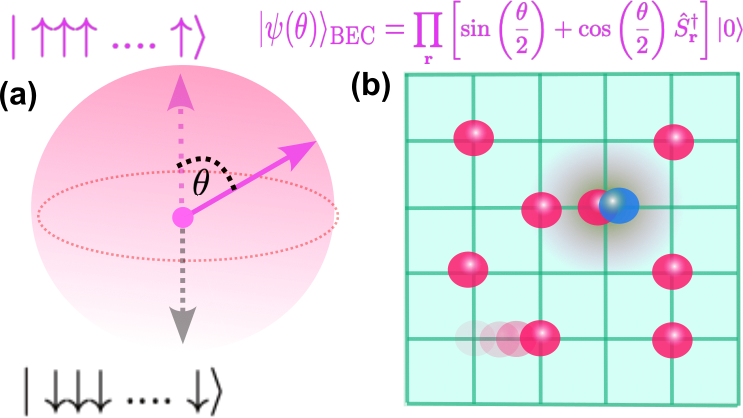}
\caption{(a) Schematic representation of the state describing a Bose-Einstein condensate (BEC) of hard-core bosons in terms of the Bloch sphere, the angle $\theta$ determines the filling factor of the BEC. (b) Cartoon of the impurity (blue ball) coupled to the BEC of hard-core bosons (pink balls). 
}
\label{FigC}
\end{figure}

Motivated by the progress with ultracold gases and the relevance in the new quantum materials, in this article, we study the strongly interacting impurity in a Bose-Einstein condensate (BEC) of hard-core bosons in the atomic context. For this purpose, we employ a variational ansatz to describe the spectral and quasiparticle properties of an impurity embedded in a gas of hard-core bosons. The character of the variational ansatz allows us to understand the coupling between the collective excitations of the hard-core gas and the impurity beyond the so-called Fr\"ohlich-like Hamiltonian, which is only valid for weak impurity-boson interactions. Our results show that the quasiparticle properties and the emergence of the polaron branches can be tuned with the underlying properties of the bath and find that for low densities, the quasiparticle properties are governed by the beyond Fr\"ohlich Hamiltonian.

Our results offer a benchmark for experiments with ultracold gases, where interactions between the bath and impurities can be tuned on demand. Furthermore, our study is motivated by recent experiments in van der Waals heterostructures, where moiré excitons and electrons can be tightly confined to a moiré superlattice and highly population Bose-Fermi mixture can be created to realize polaron physics. Our theoretical framework may provide valuable guidance for understanding these experimental systems.

% \moroni{In Section~\eqref{Model}, we outline the model, following the approach established in~\cite{bernardet2002analytical}, which characterizes hard-core bosons in terms of their elementary collective excitations. We elucidate the coupling mechanism between an impurity and the collective excitations of the gas using a  Beyond  Fr\"ohlich-like Hamiltonian. Employing a Chevy-like ansatz, we derive expressions enabling the calculation of various system properties. Moving forward to Section~\eqref{Quasiparticle properties}, we embark on numerical computations to explore the energy and residue of the polaron as functions of the coupling strength and boson density. In Section~\eqref{Spectral Features}, we delve into the analysis of the polaron's spectral function. Finally, our findings and deductions are consolidated in Section~\eqref{Conclusions}.

\section{Model}
\label{Model}
Consider a single impurity coupled to a two-dimensional gas of bosons confined in a square optical lattice of $N_s$ sites, as illustrated in Fig.~\ref{FigC}.  The Hamiltonian of the system is given by
\begin{eqnarray}\nonumber
\hat H&=&-\sum_{i=A,B}\sum_{\langle\mathbf r,\mathbf r'\rangle}\left[(t_i\hat c^\dagger_{\mathbf r,i}\hat c_{\mathbf r',i}+\text{h.c.})-\delta_{\mathbf r,\mathbf r'}\mu_{i}\hat n_{\mathbf r,i}\right]
\\ \nonumber
&+&\frac{U_{BB}}{2}\sum_{\mathbf r}\hat c^\dagger_{\mathbf r,B}\hat c^\dagger_{\mathbf r,B}\hat c_{\mathbf r,B}\hat c_{\mathbf r,B}\\
&+&U_{AB}\sum_{\mathbf r} \hat n_{\mathbf r,B}\hat n_{\mathbf r,A},\label{eq-H}
\end{eqnarray}
where the creation (annihilation) operation of the A/B atoms is denoted by $\hat c^\dagger_{\mathbf r,A/B}$ ($\hat c_{\mathbf r,A/B}$). We use index $B$ for the two-dimensional Bose gas, while index $A$ corresponds to the impurity. The tunneling coefficients are given by $t_i$, the chemical potentials by $\mu_i$, and we consider an on-site boson-boson interaction, $U_{BB}$, much larger than the tunneling, $t_B$, so we assume that the $B$ atoms can effectively be regarded as hard-core bosons. The impurity boson interaction, $U_{AB}$, is also assumed to be short-range. Finally, $\hat n_{\mathbf r,i}=\hat c^\dagger_{\mathbf r,i}\hat c_{\mathbf r,i}$ corresponds to the number operators.

Considering the majority $B$ atoms as hard-core bosons implies explicitly forbidding double occupancy of sites. To obtain the ground state of the $B$ gas and its collective excitations, we follow Refs.~\cite{bernardet2002analytical,santiago2023collective}, and we only briefly detail their proceeding. The idea is to account for the forbidden double occupancy of the $B$ atoms by mapping the bosonic field operators into spin 1/2 operators; that is, we  start by replacing $\hat c^\dagger_{\mathbf r,B}\rightarrow \hat S^\dagger_{\mathbf r},\ \text{and}\, \hat c_{\mathbf r,B}\rightarrow \hat S_{\mathbf r}$, where $\hat S^\dagger_{\mathbf r}$ ($\hat S_{\mathbf r}$) creates (annihilates) a spin $+\frac{1}{2}$ ($-\frac{1}{2}$) at location ${\mathbf r}$. For $U_{AB}=0$, the ground state of the system can be written as
 \begin{equation}\label{eq:psi_0}
   |\Phi_0\rangle= \hat c^\dagger_{\mathbf k=0,A}|\psi(\theta)\rangle_{\text{BEC}},   
 \end{equation}
this considers an impurity at the bottom of the band on top of the gas of hard-core bosons, whose many-body wave-function is given by
\begin{gather}
 |\psi(\theta)\rangle_{\text{BEC}}=\prod_{\mathbf r}\left[\sin\left(\frac{\theta}{2}\right)+\cos\left(\frac{\theta}{2}\right)\hat S^\dagger_{\mathbf r}\right]|0\rangle.
\end{gather}
Here, the angle $\theta$ is determined by a variational approach~\cite{bernardet2002analytical}: $\cos\theta=\mu_B/4t_B$, which defines the filling factor $n_B=(\cos\theta+1)/2$.

To understand the coupling between the impurity and the majority bosons, we write the Hamiltonian in Eq.~\eqref{eq-H} in terms of the collective excitations of the hard-core BEC. In turn, in momentum space, the Hamiltonian of the system is given by the sum of the following terms
\begin{equation}\label{eq-HSW-d}
    \hat H=\sum_{\mathbf k}[\omega(\mathbf k)\hat \gamma_{\mathbf k}^\dagger\hat \gamma_{\mathbf k}+\epsilon_A(\mathbf k)\hat n_{\mathbf k,A}]+\hat H_{AB},
\end{equation}
where $\hat\gamma_\mathbf{k}^\dagger$ and $\hat\gamma_{\mathbf k}$ are the creation and annihilation bosonic operators describing the collective excitations of the BEC, respectively, with  $\omega(\mathbf k)=2\sqrt{\alpha^2_{\mathbf k}-\beta^2_{\mathbf k}}$ and
\begin{subequations}
    \begin{equation}
        \alpha_{\mathbf k}=\frac{1}{2}\left[\frac{\epsilon_{\mathbf k}}{2}(\cos^2\theta+1)+4t_B\right],
    \end{equation}
    \begin{equation}
        \beta_{\mathbf k}=-\frac{1}{4}\sin^2\theta\epsilon_{\mathbf k}.
    \end{equation}
\end{subequations}
Here, $\epsilon_{\mathbf k}=-2t_B[\cos{(k_{x}a)}+\cos{(k_{y}a})]$, where $a$ is the lattice constant.  
The coupling between the impurity and the excitations of the BEC is described by $\hat H_{AB}$ as follows
\begin{gather}\label{Hamiltonian}
     \hat{H}_{AB}=\frac{U_{AB}}{2}(1+\cos{\theta})\sum_{\mathbf r} \hat n_{\mathbf r,A}\\ \nonumber
     -\frac{U_{AB}}{\sqrt{N_{s}}}\frac{\sin\theta}{2}\sum_{\mathbf{k,q}}(u_{\mathbf q}-v_{\mathbf q})(\hat{\gamma}^{\dagger}_{-\mathbf q}+\hat{\gamma}_{\mathbf q})\hat{c}^\dagger_{\mathbf k+\mathbf q,A}\hat{c}_{\mathbf k,A}\\ \nonumber
     -\frac{1}{N_{s}}U_{AB}\cos{\theta}\sum_{\mathbf {k,q,k^{'}}}\big[ u_{\mathbf k^{'}+\mathbf q}u_{\mathbf k^{'}}\hat{\gamma}^{\dagger}_{\mathbf k^{'}+\mathbf q}\hat{\gamma}_{\mathbf k^{'}}\\ \nonumber
     -u_{\mathbf k^{'}+\mathbf q}v_{\mathbf k^{'}}\hat{\gamma}^{\dagger}_{\mathbf k^{'}+\mathbf q}\hat{\gamma}^{\dagger}_{-\mathbf k^{'}}-v_{\mathbf k^{'}+\mathbf q}u_{\mathbf k^{'}}\hat{\gamma}_{\mathbf -\mathbf (\mathbf k^{'}+\mathbf q)}\hat{\gamma}_{\mathbf k^{'}}\\ \nonumber
     +v_{\mathbf k^{'}+\mathbf q}v_{\mathbf k^{'}}\hat{\gamma}_{\mathbf -(\mathbf k^{'}+\mathbf q)}\hat{\gamma}^{\dagger}_{-\mathbf k^{'}}\big] \hat{c}^\dagger_{\mathbf k-\mathbf q,A}\hat{c}_{\mathbf k,A},
\end{gather}
where the coherence factors of the BEC, $u_{\mathbf k}$ and $v_{\mathbf k}$, are defined as  
\begin{subequations}
   \begin{equation}
        u_{\mathbf k}=\sqrt{\frac{1}{2} + \frac{\alpha_{k}}{2\sqrt{\alpha_{k}^{2}-\beta_{k}^{2}}}},
   \end{equation}
    \begin{equation}
        v_{\mathbf k}=\sqrt{-\frac{1}{2} + \frac{\alpha_{k}}{2\sqrt{\alpha_{k}^{2}-\beta_{k}^{2}}}}.
    \end{equation}
\end{subequations}

The impurity-boson Hamiltonian of Eq.~\eqref{Hamiltonian} consists of three contributions: i) The mean-field term in the first line, ii) the Fr\"ohlich-like interaction where a single collective excitation can be either absorbed or emitted (second line), and iii) beyond Fr\"ohlich contributions (third to fifth lines), which involve processes with two collective modes. Notice that for $\theta\rightarrow\pi$, the interaction Hamiltonian is completely governed by the beyond Fr\"ohlich term. 

Next, we propose a Chevy-like ansatz \cite{Chevy2006,Weiran2014,Ding2022a} for the ground state as
\begin{equation}
    |\Psi\rangle = \left[\phi_{\mathbf{0}}\hat{c}^{\dagger}_{\mathbf{k}=0,A}+\sum_{\mathbf{q}}\phi_{\mathbf{q}}\hat{c}^{\dagger}_{\mathbf{q},A}\hat{\gamma}^{\dagger}_{\mathbf{-q}}\right]|\psi(\theta) \rangle_{\text{BEC}},
    \label{Chevy}
\end{equation}
with the variational parameters $\phi_{\mathbf 0}$ and $\phi_{\mathbf{q}}$, obtained from the variational principle $\delta\langle \Psi| \hat H-E|\Psi\rangle/\delta \phi_{\mathbf{0}}^{*}=0$ and  $\delta\langle \Psi| \hat H-E|\Psi\rangle/\delta \phi_{{\mathbf p}}^{*}=0$, respectively. This ansatz leads to the following set of equations:
\begin{subequations}\label{Ecs}
    \begin{equation}
        \epsilon_{MF}\phi_{\mathbf{0}}-\frac{U_{AB}}{2\sqrt{N_{s}}}\sin{\theta}\sum_{\mathbf{p}}\phi_{\mathbf{p}}(u_{\mathbf{p}}-v_{\mathbf{p}})=E \phi_{\mathbf{0}}, \label{eq1}
    \end{equation}
    \begin{gather}
        \left[\omega(\mathbf p)+\epsilon^{(A)}_{\mathbf{p}}+\epsilon_{MF}\right]\phi_{\mathbf{p}}-\frac{U_{AB}}{2\sqrt{N_{s}}}\sin{\theta}(u_{\mathbf{p}}- v_{\mathbf{p}}) \phi_{\mathbf{0}}\nonumber\\
    -\frac{1}{N_{s}}U_{AB}\cos{\theta}\sum_{\mathbf{p^{'}}}(u_{\mathbf{p}}u_{\mathbf{p^{'}}}+v_{\mathbf{p}}v_{\mathbf{p^{'}}})\phi_{\mathbf{p^{'}}}=E \phi_{\mathbf{p}},\label{Ecs2}
    \end{gather}
\end{subequations}
where $\epsilon_{MF}= n_BU_{AB}$ is the mean-field energy and 
$\epsilon^{(A)}_{\mathbf{p}}=-2t_A[\cos{(k_{x}a)}+\cos{(k_{y}a})-2]$ is the impurity dispersion. 

For the two-dimensional Bose polaron in homogeneous confinement, Chevy's ansatz in Ref.~\cite{Nakano2024} has provided a very accurate description compared to quantum Monte Carlo calculations~\cite{Ardila2020}, and it is equivalent to the non-self-consistent T-matrix approximation~\cite{cardenas2022strongly}. Furthermore, the limitations of Chevy's ansatz for weakly interacting BEC are related to the formation of clusters around the impurity~\cite{levinsen2021quantum}, clusters that, due to the bosonic nature of the atoms forming the BEC, can be arbitrarily large. 
The formation of large polaron clouds cannot be correctly captured with this ansatz. Here, however, we work in the opposite regime where these clusters are intrinsically prevented due to the hard-core nature of the bosons. Therefore, we expect this ansatz to be even more reliable with the hard-core constraint for the bath bosons.

\section{Lattice polarons}
\label{Quasiparticle properties}
We can now investigate the spectral features of the impurity coupled to the hard-core gas and the properties of the resulting lattice polaron. In particular, we perform calculations of the quasiparticle residue and the spectral function. The polaron energy is simply given by $E$ since we set the ground state energy of the BEC and the impurity to zero
The quasiparticle residue, $Z$, is determined by the squared overlap between the interacting $|\Psi\rangle$ and the non-interacting $|\Phi_0\rangle$ states, the latter corresponding to the case where the impurity sits on top of the BEC without creating correlations, see Eq.~\eqref{eq:psi_0}:
\begin{gather}
\label{residuoP}
Z=|\langle \Phi_0|\Psi\rangle|^2.    
\end{gather}
On the other hand, the spectral function of the impurity is defined by the expression
\begin{equation}
    A(\mathbf k=0,\omega)=-\frac{1}{\pi}\mathbf{Im} \bigg[ \sum_{n=0}\frac{|\langle \Phi_{0}|\Psi_n\rangle|^{2}}{\omega - E_{n}+i \eta} \bigg],
\end{equation}
where $\{|\Psi_n\rangle\}$ is the set of eigenstates obtained from the diagonalization of Eqs.~\eqref{eq1} and~\eqref{Ecs2}, and $E_n$ denotes the corresponding eigenenergies. For visibility purposes, we add a small imaginary number, $i\eta$, to the spectral function. As we focus on the zero-momentum impurity, we simply denote $A(\mathbf k=0,\omega)=A(\omega).$

\begin{figure}[h!]
\centering
\includegraphics[width=1\columnwidth]{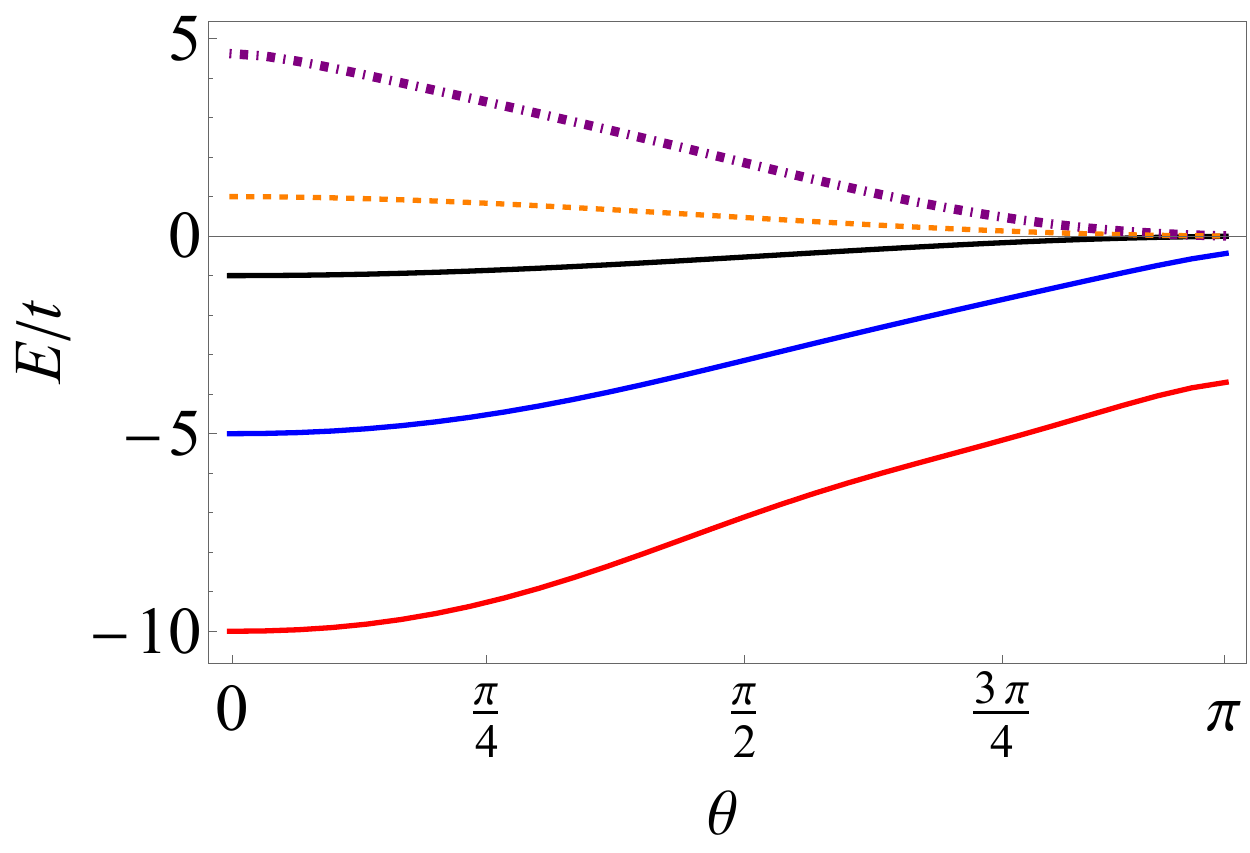} 
\caption{Zero-momentum polaron energies as a function of the angle $\theta$ for $U_{AB}=-10.0$ (red line), $U_{AB}=-5.0$ (blue line), $U_{AB}=-1.0$ (black line), $U_{AB}=1.0$ (orange line) and  $U_{AB}=5.0$ (purple line). }
\label{Energy}
\end{figure}

First, we study the  zero-momentum polaron energy, $E/t$, as a function of the angle $\theta$ (which varies the filling factor $n_{B}$) for several values of the interaction strength $U_{AB}.$ This is shown in Fig.~\ref{Energy}, where we show for attractive impurity-boson interactions: $U_{AB}=-10.0$  (red line), $U_{AB}=-5.0$ (blue line), and $U_{AB}=-1.0$ (black line) as well as for repulsive interactions $U_{AB}=1.0$ (dashed orange line) and  $U_{AB}=5.0$ (dot-dashed purple line). Unless explicitly stated, all of our simulations are performed for a squared two-dimensional lattice of $151 \times 151$ sites. We take $t_A=t_B=t.$

For attractive interactions in Fig.~\ref{Energy}, we observe for small angles, i.e., close to unit filling factors, that the polaron energy simply saturates at $U_{AB}$. 
Physically, this is very intuitive: when the bosons fill the lattice, there is no longer a continuum of scattering processes, and the energy of adding an impurity is simply given by the energy cost of putting a boson and an impurity together, which is $U_{AB}.$ 
With increasing angle, that is, decreasing the filling factor, scattering processes start to be allowed, then, the energy of the polaron decreases smoothly with the angle $\theta.$
Interestingly, we find that when $\theta\rightarrow \pi$, that is, $n_B\rightarrow 0$, the energy of the polaron does not tend to zero. This is somewhat surprising because in the limit  $n_B\rightarrow 0$, the problem can be regarded as a $B$ atom placed in an empty lattice, and one would expect that the energy of the polaron to approach zero as $n_B\rightarrow 0$. 
However, this is a partial description and it is not entirely valid in the strong coupling regime. To further understand this unexpected behavior, note that in Eq.~\eqref{Hamiltonian}, the mean-field and Fr\"ohlich terms of the Hamiltonian vanish when $\theta\rightarrow \pi$, while the term beyond Fr\"ohlich remains finite (last three lines in Eq.~\eqref{Hamiltonian}). 
This term increases with $U_{AB}$, leading to a non-zero polaron energy for $n_B\rightarrow 0$. In this limit, both the mean-field and Fr\"ohlich contributions become negligible, and the polaron state is completely dominated by the {\it beyond Fr\"ohlich} term of the Hamiltonian, this is a genuine result of strong interactions.

For repulsive interactions, observe that close to unit filling $(\theta=0)$, the energy of the polaron saturates to $E\sim U_{AB}$, as shown in Fig.~\ref{Energy}. With increasing angle (decreasing filling factor), the polaron energy smoothly decreases and tends to zero as the filling factor approaches zero. Our findings suggest that for strong coupling, the repulsive and attractive polaron behave differently in the limit $n_B\rightarrow 0$.

\begin{figure}[h]
\centering
\includegraphics[width=1\columnwidth]{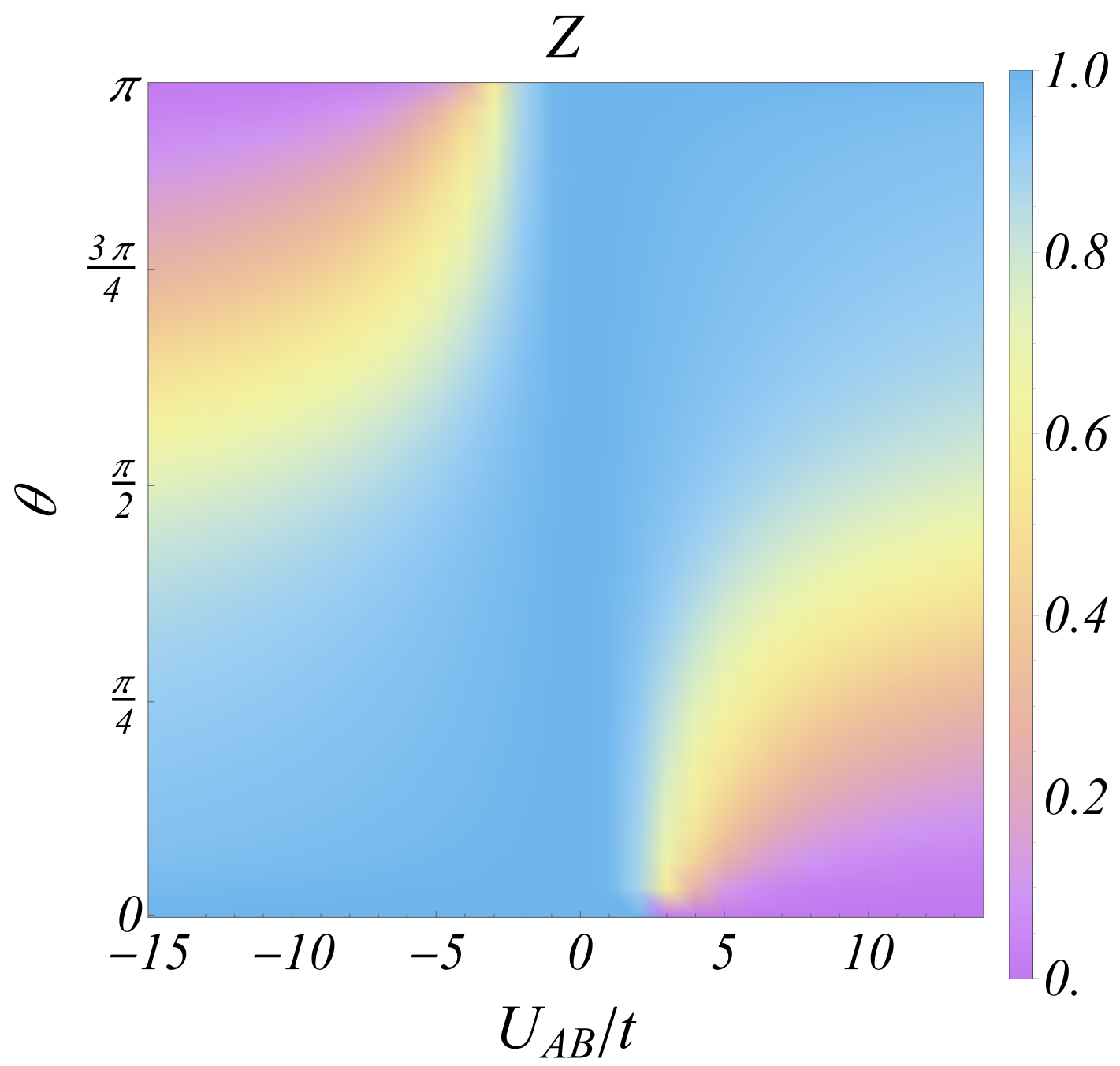} 
\caption{Quasiparticle residue, $Z$, for the zero-momentum polaron as a function of the interaction strength, $U_{AB}$, and the angle $\theta$. }
\label{Residue}
\end{figure}

To understand the quasiparticle character of the polaron states discussed above, we now turn to the polaron residue as defined in Eq.~\eqref{residuoP}. In Fig.~\ref{Residue}, we show the residue $Z$ as a function of the angle $\theta$ and the interaction strength $U_{AB}/t$.
For weak attractive interactions, we find that the polaron is well-defined with $Z\sim 1$. As the interaction strength increases, the polaron remains a well-defined quasiparticle for small angles (close to unit filling factors). In contrast, for small filling factors ($\theta \rightarrow \pi$), the residue of the polaron starts dropping to zero. That is, the states we discussed previously, where the energy of the ground-state remains non-zero in the limit $n_B\rightarrow 0$, have indeed very small residue and are consequently ill-defined quasiparticles. Note that in this same limit ($n_B\rightarrow 0$), the residue of the repulsive polaron is close to one, meaning that in this case, this branch remains as a well-defined quasiparticle, contrary to the attractive polaron. On the other hand, we obtain that close to unit filling and for strong coupling, the attractive polaron remains well-defined whereas the repulsive polaron cedes its spectral weight and is no longer a well-defined quasiparticle.
One should remark that the breakdown of the quasiparticle picture occurrs only for strong interactions. For weak impurity-boson interactions, we obtain that the quasiparticle picture holds valid for all filling factors, including the limit cases of $n_B\rightarrow 0$ and $n_B\rightarrow 1.$ 

\begin{figure}[h]
\centering
\includegraphics[width=1\columnwidth]{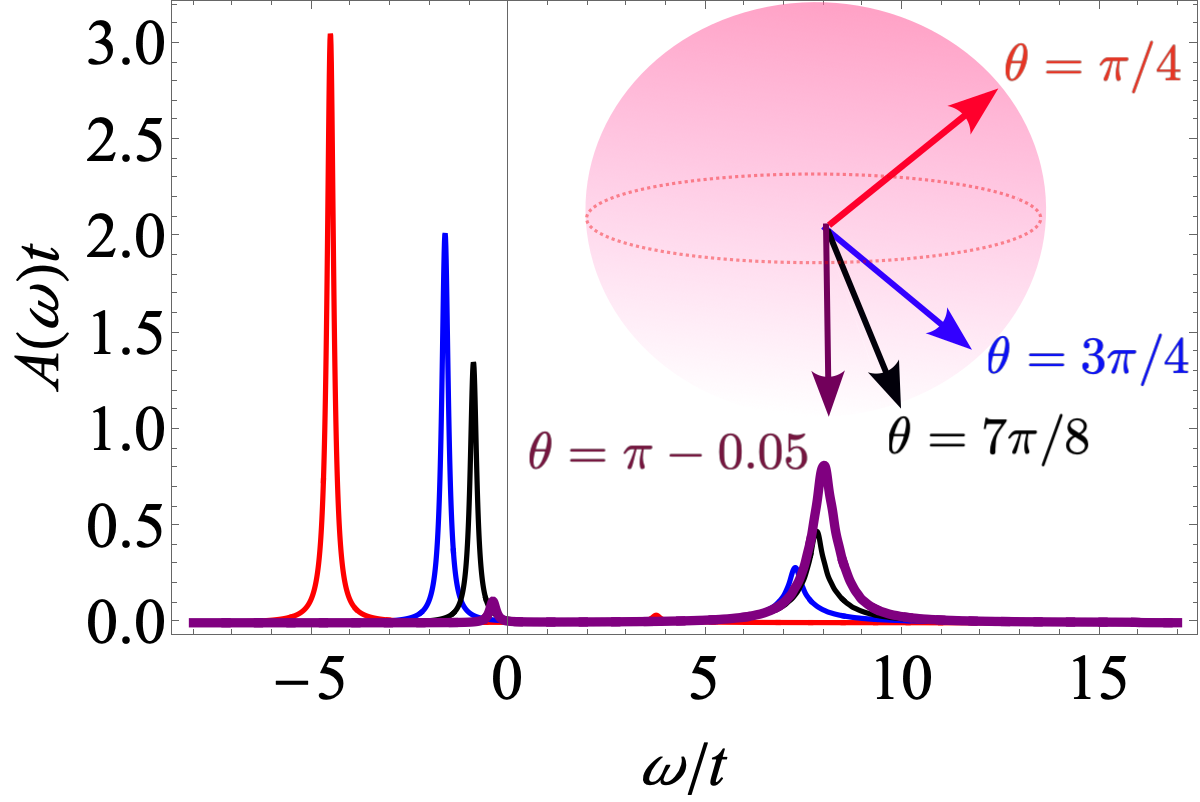}
\caption{Spectral function, $A(\omega)$, as a function of the frequency $\omega/t$ with $U_{AB}=-5$ for $\theta=\pi/4$ (red line), $\theta=3\pi/4$ (blue line), $\theta=7\pi/8$ (black line) and $\theta=\pi-0.05$ (purple line). With increasing angle (decreasing the filling factor), we observe that the polaron becomes ill-defined and cedes its spectral weight to the continuum of collective excitations.}
\label{SpectralF}
\end{figure}

Since our approach preserves the sum-rule $\int  A(\omega) d\omega=1$, if the polaron loses its residue, the spectral weight has to be distributed in high-energy excitations. To understand how the spectral weight is transferred into high-energy excitations, we plot in Fig.~\ref{SpectralF} the spectral function, $ A(\omega)$, as a function of the frequency $\omega/t$ for fixed $U_{AB}/t=-5$ and several filling factor values.

Fig.~\ref{SpectralF} shows the evolution of the polaron peak with decreasing density (increasing $\theta$) for strong impurity-boson interactions. For large filling factors (small angles), the polaron peak retains most of the spectral weight, as shown by the red line in Fig.~\ref{SpectralF}; in this case, only the quasiparticle peak at $\omega/t\approx U_{AB}=-5$ is visible in the spectral function.  
By decreasing the filling factor (increasing $\theta$), we observe a reduction of polaron peak, as indicated by the blue and black curves for $\theta=3\pi/4$ and $\theta=7\pi/8$, respectively. As the polaron peak shrinks, its spectral weight is transferred into high-energy excitations, these excitations are incoherent and consist of the continuum of excitations of our ansatz in Eq.~\eqref{Chevy}; i.e., this continuum is formed by one collective excitation of the BEC and an impurity state with opposite momentum. Close to $\theta=\pi$, the polaron peak has disappeared and the quasiparticle residue vanish, distributing its spectral weight in the continuum of excitations which become more visible (purple line).

Due to the confinement of the BEC and the impurity within an optical lattice, the continuum of excitations is bounded from above and below, and its width is given by $W=4t[2+\sqrt{2(1+\cos^{2}{\theta})}]$, which depends on $\theta$.
Note that the fading of the polaron in the strong coupling regime is a result that only the beyond Fr\"ohlich term in the Hamiltonian of Eq.~\eqref{Hamiltonian} can capture.

\section{Conclusions}
In this work, we have studied the problem of an impurity coupled to a Bose-Einstein condensate of hard-core bosons in a two-dimensional optical lattice. Employing a variational ansatz to describe the formation of the Bose polaron for strong coupling, we unveiled the quasiparticle properties of the zero-momentum impurity and have shown the interplay between the strong impurity-boson interactions and the inherent collective excitations of the hard-core bosons which tune the quasiparticle features. 

The ability to probe lattice systems site-by-site using quantum microscopy has increased interest in the study of lattice polarons, which may serve as probes to measure quantum phase transitions~\cite{colussi2023lattice}, topological and geometric features~\cite{grusdt2016interferometric,Camacho2019,Pimenov2021,pimenov2024polaron}, as well as following spatially non-equilibrium dynamics~\cite{isaule2024bound}. Entering into the strongly interacting regime may lead to the emergence of new few-body and many-body states~\cite{Ding2022a}.

Also, the recent experiments with van der Waals materials where excitons in moiré superlattices behave as hard-core bosons make our study relevant for moiré polarons, moiré polaritons, and probing strongly correlated phases with lattice polarons.

\section{Acknowledgments} We thank David Ruiz-Tijerina for valuable discussions and comments to our manuscript.
M. S. G acknowledges to Consejo Nacional de Humanidades, Ciencias y Tecnolog\'ia (CONAHCYT) for the scholarship provided.   A.C.G. acknowledges financial support from UNAM DGAPA PAPIIT Grant No. IA101923, and UNAM DGAPA PAPIME Grants No. PE101223 and No. PIIF23.
S.G.C-L acknowledges financial support from UNAM DGAPA PAPIIT Grant No.
TA100724. 

{\bf Data availability statement.-}
The data that support the findings of this study are openly available at the following URL/DOI: https://doi.org/10.5281/zenodo.10841578
\bibliography{ref}
\end{document}